\documentclass[conference]{IEEEtran}
\IEEEoverridecommandlockouts
\usepackage{cite}
\usepackage{amsmath,amssymb,amsfonts}
\usepackage{algorithmic}
\usepackage{graphicx}
\usepackage{textcomp}
\usepackage{xcolor}
\usepackage[caption=false,font=footnotesize]{subfig}
\usepackage{comment}
\usepackage{booktabs}
\usepackage{subcaption}
\usepackage{url}

\def\BibTeX{{\rm B\kern-.05em{\sc i\kern-.025em b}\kern-.08em
    T\kern-.1667em\lower.7ex\hbox{E}\kern-.125emX}}
\begin{document}

\title{Census Tract-Level Power Outage Prediction and Sensitivity Analysis During Extreme Events
%
%
\thanks{This work was partially supported by the Wayne State University Grant Advance Fund through the Office of Research. \\
* Corresponding Author}
}

\author{\IEEEauthorblockN{Antar Kumar Biswas, IEEE Student Member}
\IEEEauthorblockA{\textit{Electrical and Computer Engineering} \\
\textit{Wayne State University,}  \\
\textit{Detroit, USA}\\
hr2122@wayne.edu
}
\and
\IEEEauthorblockN{Masoud H. Nazari*, IEEE Senior Member}
\IEEEauthorblockA{\textit{Electrical and Computer Engineering} \\
\textit{Wayne State University,}  \\ \textit{Detroit, USA}\\
\textit{Electrical and Computer Engineering} \\
\textit{University of Waterloo, Waterloo, Canada}\\
masoud.nazari@wayne.edu
}
}
\maketitle

\begin{abstract}
This paper develops a two-stage hurdle model for predicting power outage occurrence and severity at the census-tract level. The proposed framework is then used to assess the sensitivity of power outage to socioeconomic, demographic, and environmental factors during extreme events.
Five heterogeneous data streams are integrated at the census tract level: 15-minute customer outage data, OpenMeteo hourly weather records, American Community Survey (ACS) socioeconomic indicators, Centers for Disease Control (CDC) social vulnerability indices (SVI), and Geographic Information System (GIS) derived vegetation coverage. 
The proposed framework is validated using a high-resolution power outage dataset covering 290 census tracts in the Detroit area over a period exceeding 14 months, with a temporal resolution of 15 minutes.
\end{abstract}

\begin{IEEEkeywords}
Extreme event, power outage prediction, social vulnerability index, meteorological variables, two-stage hurdle model.
\end{IEEEkeywords}

\section{Introduction}

Power outages impose significant economic, public health, and safety burdens on communities. The average U.S. electricity customer experienced approximately eight hours of power interruptions in 2020~\cite{eia2021}. As extreme weather events and
other disruptions increasingly challenge grid reliability,
there is growing interest in improving the resilience of power
systems \cite{jufri2019state}. Power system resilience refers to the ability of the electrical grid to anticipate, withstand, adapt to, and rapidly recover from extreme events such as hurricanes, ice storms, and extreme heat waves \cite{NAZARI2026113285, xu2025quantifying, do2023spatiotemporal, gonccalves2024extreme}.
Extreme weather events have emerged as one of the dominant drivers of large-scale power outages \cite{climatecentral}. 
Several studies have shown that wind-driven events, such as storms and hurricanes, are particularly influential in determining outage frequency and severity \cite{campbell2012weather,chang2007infrastructure,lee2022community, do2025spatiotemporal}. 
For instance, the results in  \cite{do2025spatiotemporal} indicate that outages were observed to be 9 times more during extreme heat, 30 times more prevalent during precipitation and up to 391 times more when heat and heavy precipitation occurred together. A statistical analysis of U.S. power system reliability found that 5\% increase in annual average wind speed is associated with approximately 14\% increase in the frequency of power interruptions\cite{larsen2016severe}. Additionally, a 10\% increase in lightning strikes correlates with about 2\% increase in reliability events. 

In our earlier works \cite{NAZARI2026113285, biswas2026data}, we presented a comprehensive review of community-centric power system resilience, highlighting the importance of integrating technical, environmental, and socioeconomic factors into resilience assessments. The findings demonstrated that SVI plays a critical role in determining community resilience and should therefore be explicitly incorporated into resilience analysis and planning frameworks.
%
In \cite{wang2024deep}, it was demonstrated that incorporating weather, socioeconomic, and power infrastructure variables enhances census tract-level outage prediction performance. Furthermore, \cite{ganz2023socioeconomic} showed that a one-decile increase in the socioeconomic status component of the SVI was associated with a 6.1\% increase in expected outage duration. For instance, during the 2021 Texas winter storm, about 4.5 million customers lost power  \cite{lee2022community}. Tracts with higher shares of low-income and minority populations experienced greater outage impacts and longer recovery times. 

Collectively, these studies highlight
the need for analytical frameworks that capture the influence
of both dynamic weather conditions and static community
characteristics on power system resilience. Prior studies primarily analyze the relationship between outages, weather conditions, and socioeconomic factors. However, these variables are typically incorporated within a
single outage model rather than being examined separately as
dynamic and static contributors to outage behavior. Additionally, most analyses are conducted at relatively large scale, such as states or counties, which may obscure localized variations in infrastructure conditions and social vulnerability. 

This paper addresses these limitations through a unified
two-stage prediction framework applied at the census tract
level in Detroit, Michigan. The proposed framework
separately models outage occurrence and outage severity,
allowing the influence of dynamic weather conditions and
static community characteristics to be evaluated more
effectively. The main contributions of
this work are:
\begin{enumerate}
  \item A two-stage census tract-level outage prediction framework
        that separately models outage occurrence and outage severity.

  \item A multi-source data integration framework combining utility
        outage records, meteorological observations, socioeconomic
        indicators, social vulnerability metrics, and vegetation
        coverage to characterize outage risk at high spatial resolution.

  \item An interpretable statistical modeling approach validated using
     leave-one-event-out (LOEO) cross-validation across four major storm
        events, providing tract-level estimates of both outage
        probability and outage magnitude.
\end{enumerate}

The remainder of this paper is organized as follows. Section II describes the data sources used in this study. 
Section III presents the methodology, including spatial data integration and statistical analysis techniques. Section IV discusses the results of the correlation and predictive analyses. Finally, Section V concludes the paper.

\section{Data Source}

This study integrates five heterogeneous datasets, all linked
at the census tract level using the 11-digit GEOID, which uniquely
identifies each census tract through a combination of state,
county, and tract codes.
A summary of the integrated data sources is provided in
Table~\ref{tab:datasources}.

\begin{table}[!t]
\caption{Summary of Data Sources.}
\label{tab:datasources}
\centering
\footnotesize
\setlength{\tabcolsep}{4pt}
\begin{tabular}{p{1.3cm}p{1.6cm}p{1.4cm}p{2.0cm}}
\toprule
\textbf{Source} & \textbf{Dataset} & \textbf{Coverage} & \textbf{Role} \\
\midrule
DTE Energy    & 15-min outage GeoJSON & 290 Detroit tracts & Outage target \\
Open-Meteo    & Hourly weather & Census tract level & Weather features \\
ACS 2023      & 5-year estimates & 290 tracts & Socioeconomic \\
CDC SVI 2022  & Vulnerability index & Census tract level & Social vulnerability \\
GIS canopy    & Vegetation coverage\% & Census tract level & Environmental data \\
\bottomrule
\end{tabular}
\end{table}

\subsection{Power Outage Data}
\begin{figure}[!t]
\centering
\includegraphics[width=0.9\linewidth]{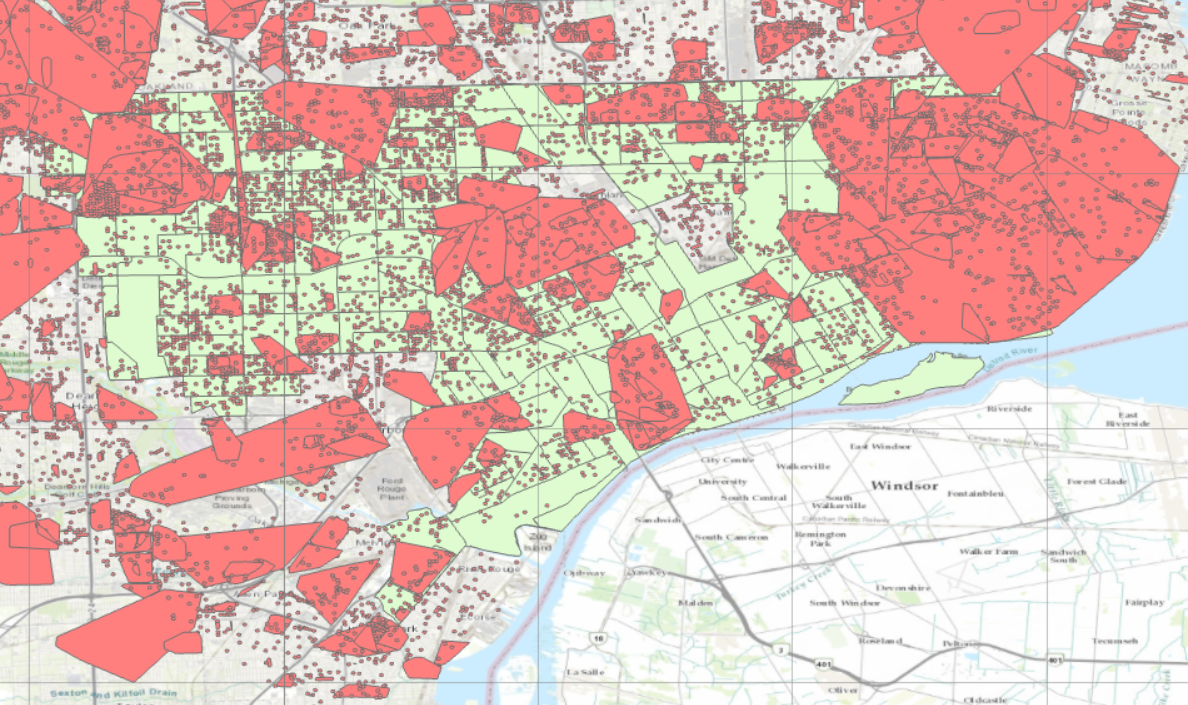}
\caption{Spatial representation of DTE Energy outage polygons (red)
overlaid on Detroit census tract boundaries. The geometric
intersection between outage polygons and census tracts is used
to allocate affected customers.}
\label{fig:outage}
\end{figure}

Power outage information is obtained from the DTE Energy outage map, which provides time-stamped outage incidents along with estimates of affected customers. The dataset is updated every 15 minutes. The dataset contains approximately 14 months of records spanning 2023–2025. These records are spatially linked to Detroit census tracts to quantify outage burden and frequency. \footnote{https://outagemap.serv.dteenergy.com/GISRest/services/OMP}
Fig. \ref{fig:outage} illustrates the geographic distribution of outage layer within Detroit census tracts.

\subsection{Socioeconomic Data}

Socioeconomic and demographic variables are derived from the U.S. Census Bureau’s ACS 5-Year Estimates \footnote{https://api.census.gov/data/2021/acs/acs5/}. Variables include median household income, poverty rate, internet access, and housing age distribution to evaluate social vulnerability and resilience. The SVI developed by the Centers for Disease Control and Prevention (CDC) and the Agency for Toxic Substances and Disease Registry (ATSDR), is incorporated to capture broader dimensions of community vulnerability. The SVI is a composite metric derived from socioeconomic, demographic, household composition, housing, and transportation characteristics. Higher SVI values indicate communities that are more vulnerable and may face greater challenges in preparing for, responding to, and recovering from extreme weather events, power outages, and other disruptive hazards.


\subsection{Weather Data}

Meteorological data are collected from the Open-Meteo platform, which provides high-resolution hourly weather variables such as temperature, wind gust, and precipitation. These weather features are integrated with outage and socioeconomic data to assess the influence of environmental conditions on outage occurrence.

\subsection{Environmental Data}

\begin{figure}[!t]
\centering
\includegraphics[width=1\linewidth]{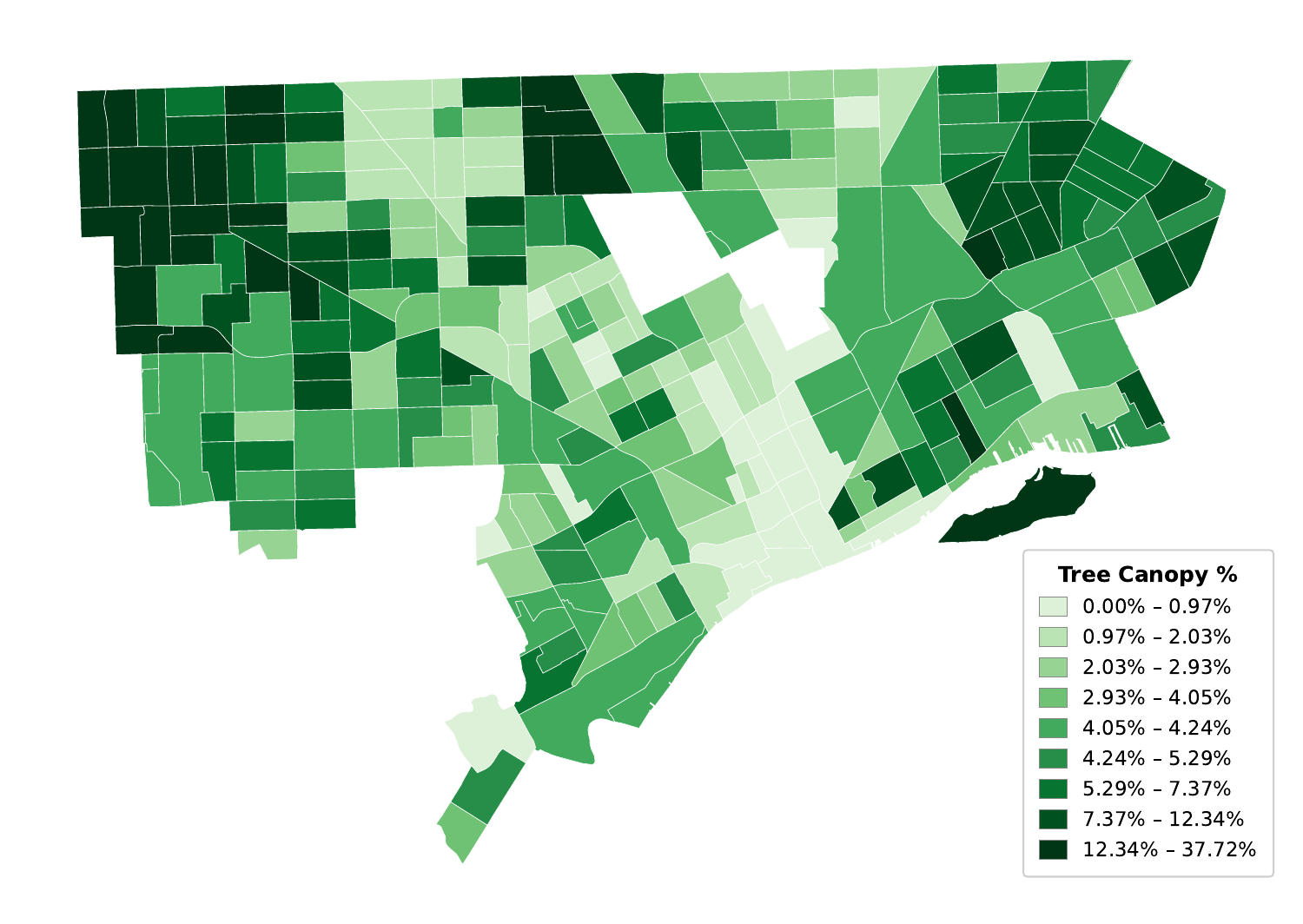}
\caption{Spatial distribution of mean vegetation density across census tracts in Detroit and surrounding areas. Darker green shades indicate higher vegetation coverage, while lighter shades represent lower density.}
\label{fig:tree_percentage}
\end{figure}

The dataset provides spatial information on vegetation coverage across the study area. Census tract–level mean vegetation density is obtained from publicly available GIS land-cover datasets and spatially joined with the census tract boundary shapefile. The vegetation data were processed and spatially joined with census tract boundaries. The variable represents the proportion of land area within each census tract covered by tree vegetation. Tree contact with power lines can increase fault frequency, prolong restoration time, and reduce overall service continuity. The spatial distribution of vegetation percentage across Detroit census tracts is illustrated in Fig. \ref{fig:tree_percentage}. 

\section{Methodology}

\subsection{Data Processing}

DTE outage polygons often cross census tract boundaries. For each
15-minute snapshot, the geometric intersection between each outage polygon
and each census tract is computed for accurate area calculations. Customers are allocated
proportionally to the fractional area overlap:
\begin{equation}
  C_{k}^{\,t} = \texttt{NUM\_CUST}_{p}
                \;\times\;
                \frac{A_{pk}}{A_p}
  \label{eq:allocation}
\end{equation}
where $C_{k}^{\,t}$ is the customers allocated to tract $k$ from polygon
$p$ at time~$t$, $A_{pk}$ is the intersection area, and $A_p$ is the total
polygon area. When multiple polygons overlap the same tract at the same
timestamp, their contributions are summed. 

After constructing
the tract-level outage dataset, storm events are identified using DTE's operational \texttt{STORM\_MODE}
flag, which is activated system-wide during major weather incidents. Each
snapshot file is scanned for any polygon carrying
\texttt{STORM\_MODE = "STORM"}. Of 30,238 snapshots, 1,626 (5.4\%) carry
this flag. Consecutive storm-mode snapshots within a 3-hour gap are merged
into a single event, yielding four distinct storm events in 2024, all
falling within the study window. For each event, the outage magnitude per
tract is taken as the peak (maximum) customers out across all within-event
snapshots:
\begin{equation}
  C_{ij}^{\mathrm{peak}}
  = \max_{t \,\in\, \text{event}\,j}\!\bigl\{C_{i}^{\,t}\bigr\}.
  \label{eq:peak}
\end{equation}
Table~\ref{tab:events} summarizes the four identified events.

\begin{table}[!t]
\caption{Identified Storm Events — Detroit, Michigan.}
\label{tab:events}
\centering
\footnotesize
\setlength{\tabcolsep}{3pt}
\begin{tabular}{clp{1.6cm}cc}
\toprule
\textbf{ID} & \textbf{Date} & \textbf{Category} &
\textbf{Tracts Affected} & \textbf{Min Temp ($^\circ$C)} \\
\midrule
E1 & Jan 13--20 & Winter Cold Event & 162 (55.9\%) & $-$21.0 \\
E2 & Jun 5--9   & Tornado           &  79 (27.2\%) &    14.2 \\
E3 & Jun 23--25 & Wind/Storm        &  36 (12.4\%) &    17.8 \\
E4 & Jun 27--29 & Wind/Storm        &  41 (14.1\%) &    18.1 \\
\bottomrule
\end{tabular}
\end{table}

For each storm event, three weather features
are aggregated per census tract: minimum
temperature (\textit{temp\_min}), total precipitation (\textit{precip\_sum}), and maximum wind gust
(\textit{gust\_max}).
The cross-combination of 290 modeled census tracts and four
storm events produced a total of 1,160 tract-event
observations, of which 318 (27.4\%) correspond to observed outage cases. The binary
target variable is defined as:
\begin{equation}
  Y_{ij} = \mathbf{1}\!\left[C_{ij}^{\mathrm{peak}} > 0\right],
  \label{eq:target}
\end{equation}
where $Y_{ij}=1$ indicates that census tract $i$ experienced an outage during storm event $j$, and $Y_{ij}=0$ otherwise.

\subsection{Two-Stage Statistical Model}

The two-stage model decomposes the expected customers outage per
tract-event as:
\begin{equation}
\begin{aligned}
\mathbb{E}[C_{ij}]
&=
\underbrace{
P(Y_{ij}=1 \mid \mathbf{W}_{ij}, \mathbf{X}_{i}, \mathbf{S}_{i})
}_{\text{Stage A}}
\\[4pt]
&\quad \times
\underbrace{
\mathbb{E}[C_{ij} \mid Y_{ij}=1,
\mathbf{W}_{ij}, \mathbf{X}_{i}, \mathbf{S}_{i}]
}_{\text{Stage B}},
\end{aligned}
\label{eq:twostage}
\end{equation}
where $\mathbf{W}_{ij}$ represents weather-related dynamic
variables, $\mathbf{X}_{i}$ represents infrastructure variable, and $\mathbf{S}_{i}$ denotes
static socio-economic variables that remain relatively constant over the study period for census tract $i$.

In Stage A, logistic regression is performed to model the probability of the
outage as:
\begin{equation}
\mathrm{logit}\!\left(P(Y_{ij}=1)\right)
=
\beta_0
+
\boldsymbol{\beta}_w^{\top}\mathbf{W}_{ij}
+
\boldsymbol{\beta}_x^{\top}\mathbf{X}_{i}
+
\boldsymbol{\beta}_s^{\top}\mathbf{S}_{i},
\label{eq:stageA}
\end{equation}
where coefficient vectors
$\boldsymbol{\beta}_w$, $\boldsymbol{\beta}_x$, and
$\boldsymbol{\beta}_s$ quantify the contributions of the
corresponding feature groups. Each coefficient $\beta$ represents the change in log-odds
of outage occurrence per one-standard-deviation increase in
the predictor, while the exponentiated coefficient
$\exp(\beta)$ corresponds to the odds ratio (OR).

In Stage B, negative binomial regression is performed to model the count of
customers out:
\begin{equation}
\begin{aligned}
\log\!\left(\mathbb{E}[C_{ij} \mid Y_{ij}=1]\right)
&=
\log(N_{ij})
+
\gamma_0
+
\boldsymbol{\gamma}_w^{\top}\mathbf{W}_{ij}
\\
&\quad
+
\boldsymbol{\gamma}_x^{\top}\mathbf{X}_{i}
+
\boldsymbol{\gamma}_s^{\top}\mathbf{S}_{i}.
\end{aligned}
\label{eq:stageB}
\end{equation}

Here coefficient
vectors $\boldsymbol{\gamma}_w$, $\boldsymbol{\gamma}_x$,
and $\boldsymbol{\gamma}_s$ quantify the contributions of the
corresponding feature groups. Furthermore, $N_{ij}$ denotes
the total number of households associated with tract $i$
during storm event $j$. Exponentiated coefficients $\exp(\gamma)$ represent rate ratios. 

Five tract-level static features are used: poverty rate, fraction of pre-1950 homes, vegetation coverage (canopy pct), SVI, 
and total households. Three event-level dynamic features are used: peak wind gust, total precipitation, and minimum temperature. All predictors are standardized (z-scored) prior to model fitting so that coefficients represent the effect of a one-standard-deviation (1-SD) increase and are directly comparable across variables. Table~\ref{tab:sd_features} reports the mean and standard deviation of each feature.

\begin{table}[!t]
\centering
\caption{Descriptive Statistics of Model Features (1-SD Interpretation).}
\label{tab:sd_features}
\begin{tabular}{lccc}
\toprule
\textbf{Feature} & \textbf{Mean} & \textbf{SD (= 1-SD)} & \textbf{Type} \\
\midrule
Poverty rate (\%)          & 31.49 & 13.64 & Static \\
Pre-1950 homes (\%)        & 16.39 & 12.74 & Static \\
Vegetation coverage (\%)           &  5.97 &  6.03 & Static \\
SVI composite (0--1)       &  0.75 &  0.21 & Static \\
Log(total households)      &  6.80 &  0.49 & Static \\
Peak wind gust (mph)       & 60.05 &  9.68 & Dynamic \\
Total precipitation (mm)   & 10.29 &  4.20 & Dynamic \\
Min temperature (\textdegree C) &  6.76 & 16.04 & Dynamic \\
\bottomrule
\end{tabular}
\end{table}

With only four storm events available,
LOEO cross-validation is used. In each fold, one storm
event is reserved for testing, while the remaining three
events are used for training. This process is repeated four
times so that each storm event serves once as the test set. 

\subsection{Performance Metrics}

The Odds Ratio (OR) and Rate Ratio (RR) metrics used to evaluate the outage prediction model are summarized below.

OR quantifies how the odds of outage occurrence
change for 1-SD increase in a
predictor. It is obtained by exponentiating the corresponding
Stage A logistic regression coefficient:
\begin{equation}
\mathrm{OR}_k = e^{\beta_k} = \frac{P(Y=1\mid x_k{+}1)/P(Y=0\mid x_k{+}1)}{P(Y=1\mid x_k)/P(Y=0\mid x_k)},
\end{equation}
where $x_k$ denotes the current value of the predictor $k$, and $x_k+1$ represents the value of predictor $k$ after
1-SD increase. OR\,=\,1.0 means no effect. OR\,=\,1.5 means the odds of outage are 50\% higher per 1-SD increase. OR\,=\,0.37 means the odds are reduced to 37\% of their original value — equivalently, $1/0.37 = 2.7\times$ more likely when the predictor decreases by 1-SD (e.g., colder temperature).

RR measures how the expected number of affected
customer changes, conditional on an outage already occurring.
It is obtained by exponentiating the corresponding Stage B
negative binomial regression coefficient:
\begin{equation}
\mathrm{RR}_k = e^{\gamma_k} = \frac{\mathbb{E}[C\mid Y=1,\,x_k{+}1]}{\mathbb{E}[C\mid Y=1,\,x_k]}.
\end{equation}

RR\,=\,1.0 means no effect on severity. RR\,=\,1.13 means 13\% more customers are interrupted per 1-SD increase. RR\,=\,0.77 means 23\% fewer customers are interrupted ($1/0.77 = 1.30\times$ more when the predictor decreases by 1-SD).

\section{Results}

\subsection{Stage A: Outage Occurrence Model}

Stage~A uses logistic regression to estimate the probability that a census
tract experiences at least one customer outage during a given storm event. Model coefficients are interpreted using OR. 

\begin{figure}[!t]
\centering
\includegraphics[width=\linewidth]{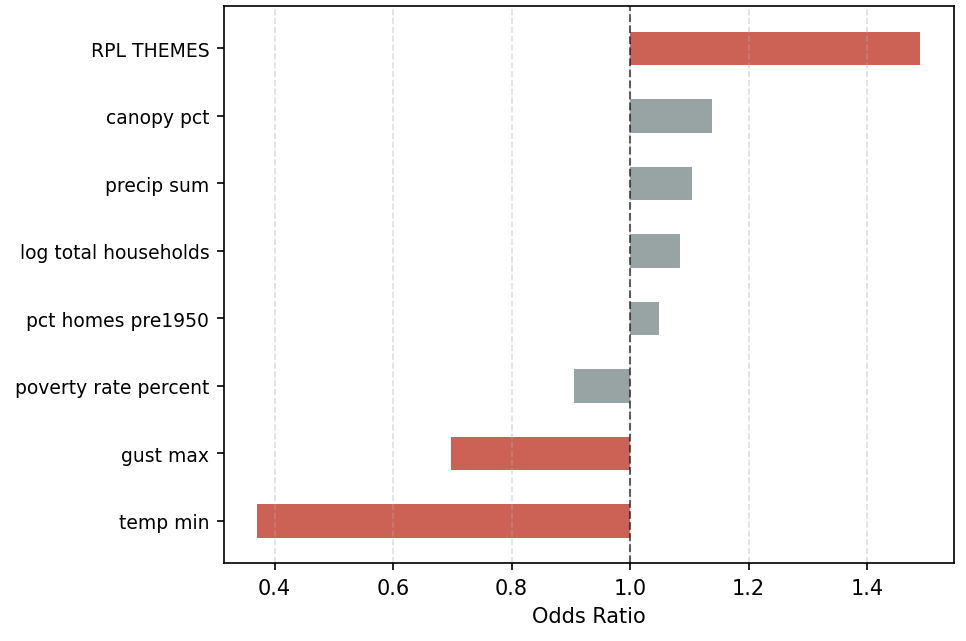}
\caption{Stage A coefficient plots. Stage~A odds
ratios (logistic regression). All predictors scaled to unit SD.} 
\label{fig:coefs}
\end{figure}

Minimum temperature, with an OR of 0.37, is a strong
predictor in Stage~A, where each 16\textdegree C decrease in
minimum temperature increases the odds of any outage by a
factor of $1/0.37 = 2.7\times$.
This result is primarily driven by the contrast between Event~E1 (winter
storm, min~temp $-$21$^\circ$C) and the three
summer convective events (min temps 14–18$^\circ$C). Temperature thus captures both the direct thermal stress on
overhead distribution equipment and the systemic contrast between winter
and summer storm types in the dataset.

An OR of 1.49 for social vulnerability indicates that
tracts with higher SVI values exhibit greater
outage likelihood, corresponding to a 49\% increase in outage
odds per 0.21 increase in SVI. More vulnerable communities may have aging or less resilient distribution infrastructure, higher exposure to environmental stressors, and fewer mitigation resources, all of which increase susceptibility to outage initiation during extreme events.

Wind gust produces an OR of 0.70, indicating a
negative association with outage occurrence after accounting
for temperature effects. This behavior is influenced by the
event composition of the dataset, where the winter storm event
experienced the highest outage impact despite lower average
gust levels than the summer storm events. As a result, higher
gust intensity alone does not independently correspond to
increased outage probability within this four-event sample.

Vegetation coverage yields an OR of 1.14, suggesting
that outage likelihood increases in tracts with
denser vegetation. This finding is consistent with
vegetation-related outage mechanisms, particularly tree-line
contact during severe weather conditions.

The results indicate that outage occurrence is primarily driven by temperature and social vulnerability, with vegetation coverage playing a secondary role. Poverty rate, pre-1950 housing fraction, and total precipitation are not statistically significant, indicating no reliable association with outage occurrence in this dataset.

The model correctly classifies 162 outage
tracts and 714 non-outage tracts, resulting in an overall
classification accuracy of 75.5\%. These results demonstrate
that the proposed framework provides reasonably reliable
outage classification performance across different storm
events.

\subsection{Stage B: Outage Severity Model}

Stage~B is estimated on the 290 tract-event observations with confirmed outages. The objective is to
estimate outage severity, measured as the peak number of
customers affected.

An RR of 0.77 for minimum temperature indicates that
colder events are associated with greater outage severity,
corresponding to approximately 23\% more customers affected
per household for 1-SD decrease in minimum temperature. This relationship likely
reflects the increased infrastructure stress and prolonged
impact associated with sustained freezing conditions.

Vegetation coverage produces an RR of 1.13, indicating
that outage severity increases by approximately 13\% per 1-SD
increase in canopy density among outage-affected tracts. This
result suggests that vegetation-related failures, particularly
tree-line contact during wind and ice events, remain a
contributor to outage propagation.


Table \ref{tab:stageB} shows that minimum temperature and vegetation coverage exhibit significant
associations with outage severity while
wind gust, poverty rate, SVI and pre-1950 housing percentage do
not demonstrate statistically significant effects within the
analyzed dataset.

\begin{table}[!t]
\caption{Stage B — Negative Binomial Regression Key Results (per 1 SD).}
\label{tab:stageB}
\centering
\footnotesize
\setlength{\tabcolsep}{5pt}
\begin{tabular}{lccl}
\toprule
\textbf{Feature}          & \textbf{RR}   \\
\midrule
temp\_min                 & 0.77            \\
canopy\_pct (vegetation)  & 1.13            \\
RPL THEMES (SVI)         &         not significant             \\
gust\_max, poverty\_rate, pct\_pre1950 & \multicolumn{3}{c}{not significant} \\

\bottomrule
\end{tabular}
\end{table}

\subsection{Combined Model Performance}

\begin{figure}[htbp]
    \centering
    \includegraphics[width=\columnwidth]{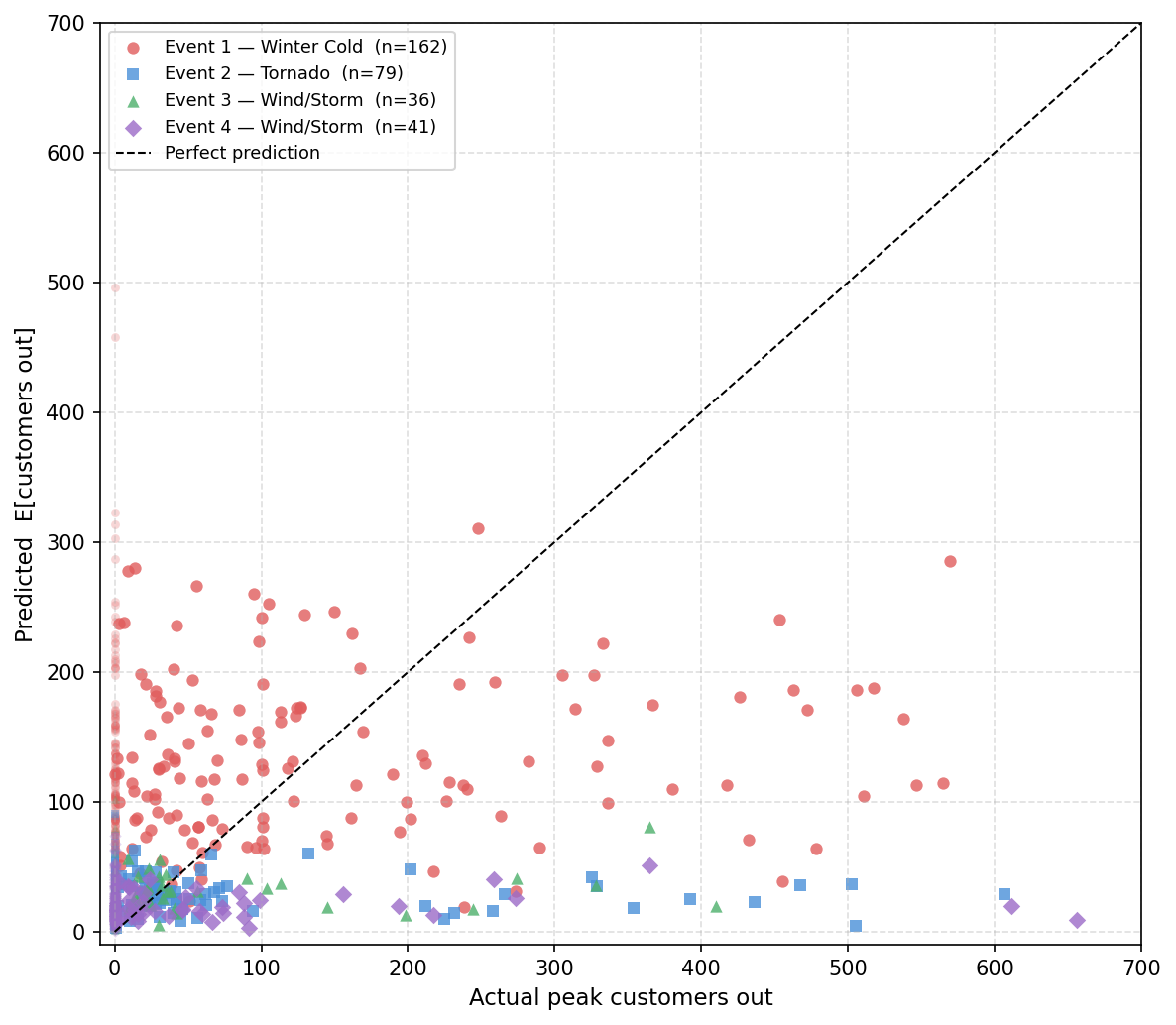}
    \caption{Combined two-stage model: predicted versus actual peak customers
    out for all census tracts.}
    \label{fig:combined_fit}
\end{figure}

The combined model integrates the two-stage hurdle framework by computing
the unconditional expected number of customers without power for every
census tract and storm event as:

\begin{equation}
    \hat{C}_{ij} = \hat{P}(Y_{ij}=1)\times\hat{E}[C_{ij}\mid Y_{ij}=1].
\end{equation}
Table ~\ref{tab:combined_performance} reports the combined model performance metrics,
and the predicted versus actual customer counts are shown in
Fig. ~\ref{fig:combined_fit}.

For each tract and storm event, the target variable represents the peak customer outage. 
The model achieves the normalized mean absolute error of 41.2\%. 

The coefficient of determination ($R^2 = 0.116$) indicates that the model explains 11.6\% of the total variance in outage magnitudes across all tract--event observations. The Pearson correlation coefficient ($r = 0.349$) demonstrates a statistically significant positive association between predicted and observed outage values, indicating that the model captures meaningful spatial and event-level outage patterns. In practice, the framework is more effective for identifying relative outage risk and prioritizing vulnerable tracts than for exact point prediction of outage magnitude. The results therefore support the use of the model as a planning and operational screening tool for extreme-event preparedness, where identifying higher-risk locations is often more critical than predicting the exact number of affected customers.

\begin{table}[htbp]
    \centering
    \caption{Combined Two-Stage Model Performance.}
    \label{tab:combined_performance}
    \normalsize
        \begin{tabular}{lc}
            \toprule
            \textbf{Metric} & \textbf{Value} \\
            \midrule
            MAE (customers)          & 67 \\
            nMAE (\% of mean actual)  & 41.2\% \\
            $R^2$                   & 0.116 \\
            Pearson $r$             & 0.349   \\
            \bottomrule
        \end{tabular}

\end{table}

\begin{figure}[htbp]
    \centering
    \includegraphics[width=\columnwidth]{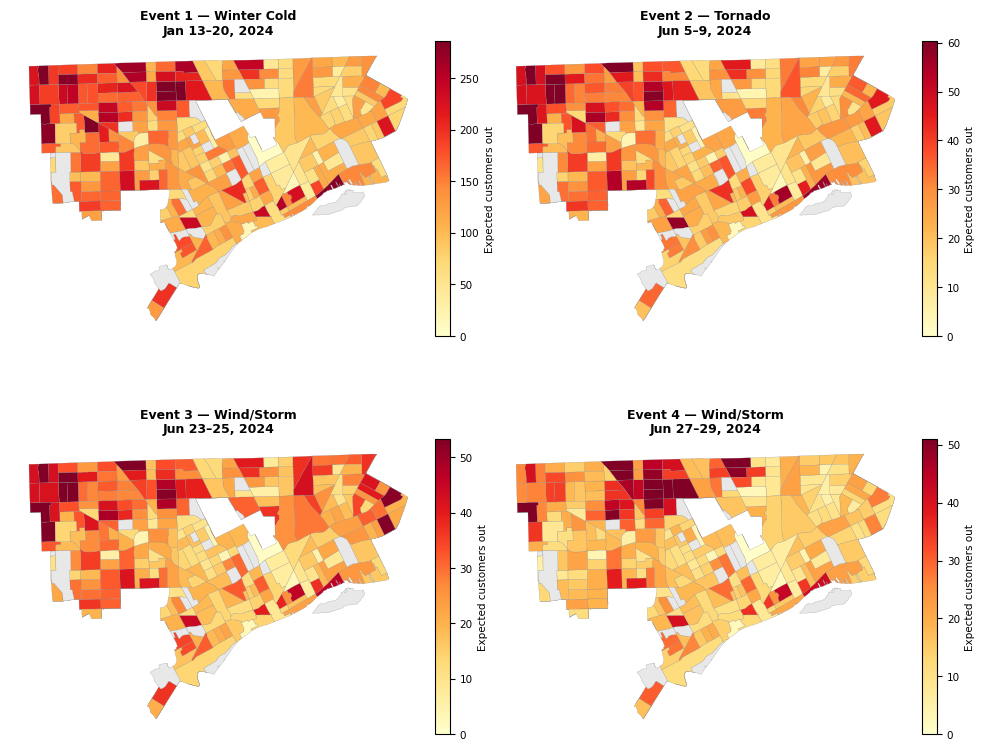}
    \caption{Spatial distribution of predicted outage customers, across Detroit census tracts for four major 2024 storm events: (Event 1) January winter storm, (Event 2) June tornado storm, and (Events 3 and 4) June windstorms.}
    \label{fig:Events}
\end{figure}

Fig. \ref{fig:Events} illustrates the predicted expected outage customers, across Detroit census tracts for the four major 2024 storm events. Two dominant spatial patterns are observed. First, strong event-dependent severity differences emerge across storm types. The January winter cold event produces substantially larger predicted outages, with several tracts exceeding 250 expected customers out, whereas the summer tornado and windstorm events generally remain below 50–60 customers per tract. This reflects the stronger county-wide outage activation associated with extreme cold conditions. 

Second, a persistent structural spatial pattern appears across all four events. Western and southwestern suburban tracts consistently exhibit the highest predicted outage levels regardless of event type. 
These suburban regions also contain greater residential vegetation coverage, which provides an additional amplification effect during severe weather conditions.

\section{Conclusions and Discussions}

This paper presented a two-stage hurdle modeling framework
for census tract-level outage prediction integrating dynamic weather variables with static
socio-economic, environmental, and infrastructure-related
features across four major storm events. The results based on the 14-month Detroit dataset spanning 2023–2025 demonstrate that weather-related factors are the dominant drivers of power outage occurrence. In particular, minimum
temperature emerged as a strong predictor in outage occurrence.  
%
The results further indicate that SVI also influences outage behavior.
Higher SVI values are associated with increased outage
probability, suggesting that socially vulnerable communities
may experience reduced infrastructure resilience and lower
levels of system investment. 

The proposed framework demonstrates that combining
dynamic meteorological information with static socioeconomic
and environmental indicators can provide meaningful insight
into both outage likelihood and outage magnitude at high
spatial resolution. The model achieved an overall outage
classification accuracy of 75.5\% under LOEO
cross-validation, demonstrating reasonably reliable predictive
performance across different storm events.

The sensitivity analysis highlights different drivers of outage
occurrence and outage severity. For instance, outage occurrence
is strongly influenced by minimum temperature and social
vulnerability. A 16\textdegree C decrease in minimum temperature
increases outage odds by approximately 2.7 times, while 1-SD increase in SVI increases outage odds
by 49\%. In addition, outage severity is associated
with minimum temperature and vegetation coverage. 
Colder
conditions lead to larger outages, while higher tree canopy
coverage is associated with more customers affected once an
outage occurs. 
These results suggest that measures such as strengthening infrastructure in vulnerable
communities and targeted vegetation management may help
reduce outage impacts during future storm events.
The proposed framework can also support utility operators in
identifying geographically vulnerable regions prior to
extreme weather events.

Despite these promising results, limitations should be
acknowledged. The analysis is based on only four major storm
events within a single year, which limits the statistical
robustness and generalizability of the estimated predictor
relationships. 
The area-proportional outage allocation
method may also introduce uncertainty because customer
density is unlikely to be uniformly distributed within outage
polygons.

Future work will focus on extending the framework to
multi-year outage datasets and additional geographic regions,
incorporating feeder and transformer-level infrastructure
information, and exploring advanced spatial-temporal learning
approaches to better capture localized outage propagation
patterns. Integrating restoration-time and operational utility
datasets may further improve the applicability of the proposed
framework for real-world resilience planning and outage
mitigation.


\end{document}